\documentclass[aps,pre,reprint,groupedaddress,showpacs]{revtex4-1}
\usepackage{graphicx}
\usepackage{dcolumn}
\usepackage{bm}
\usepackage{subfigure}
\usepackage{float}

\begin{document}


\title{Incomplete Phase-Space Method to Reveal Time Delay From Scalar Time-series}


\author{Shengli Zhu}
\author{Lu Gan}
\thanks{ganlu@uestc.edu.cn}
\affiliation{Center for Cyber Security, School of Electronic Engineering, University of Electronic Science and Technology of China, Chengdu 611731, China}


\begin{abstract}
A computationally quick and conceptually simple method to recover time delay of the chaotic system from scalar time series is developed in this paper. We show that the orbits in the incomplete two-dimensional reconstructed phase-space will show local clustering phenomenon after the component permutation procedure proposed in this work. We find that information captured by the incomplete two-dimensional reconstructed phase-space, is related to the time delay ${\tau _0}$ present in the system,
and will be transferred to the permutation component by the procedure of component permutation. We then propose the segmented mean-variance (SMV) from the permutation component to identify the time delay ${\tau _0}$ of the system.
The proposed SMV shows clear maximum when the embedding delay $\tau $ of the incomplete reconstruction matches the time delay ${\tau _0}$ of the chaotic system. Numerical data generated by a time-delay system based on the Mackey-Glass equation operating in the chaotic regime are used to illustrate the effectiveness of the proposed SMV. Experimental results show that the proposed SMV is robust to additive observational noise and is able to recover the time delay of the chaotic system even though the amount of data is relatively small and the feedback strength is weak.
Moreover, the time complexity of the proposed method is quite low.
\end{abstract}
\maketitle
\section{\label{sec:level1}Introduction}
Delay phenomena, which are due to the finite signal propagation speed or the memory effects, are ubiquitous in various systems including nonlinear optics\cite{ikeda1979multiple,lang1980external}, biology\cite{mackey1977oscillation,longtin1990noise}, chemistry\cite{epstein1992delay,roussel1996use} and climatology\cite{tziperman1994nino,clarke1998dynamics}. It is found that even a very simple time-delay chaotic system can produce highly complex dynamics with a lot of degree of freedom\cite{fischer1994high}, which makes such systems very attractive. We can find a lot of relevant applications based on the delay phenomena in nonlinear optics, for example, the chaotic radar\cite{lin2004diverse} and lidar\cite{lin2004chaotic}, the optical chaos encryption\cite{vicente2005analysis}, rainbow refractometry\cite{peil2006rainbow}, and ultrahigh-speed physical random number generation\cite{murphy2008chaotic}. The time delay is important for chaos communication since the dynamics of such delayed chaotic systems can be identified and modeled once their time delay is recovered\cite{hegger1998identifying,zhou1999extracting}. Consequently, the identification of time delay present in chaos communication systems would weaken their security and confidentiality\cite{udaltsov2005time,soriano2009security}. Besides, it is necessary to determine whether there are time delays present in the scalar time series if one wants to develop suitable models for simulation and forecasting purposes.

For the reasons aforementioned, it is very necessary to study the time delay signature present in the chaotic system. However, the great challenge is that the corresponding underlying equations or even the relevant governing mechanisms are often unknown and the time series which is always contaminated by noise is insufficient sometimes in the study of nonlinear dynamical systems.
There were a lot of approaches proposed to recover the time delay ${\tau _0}$ of the system from recorded time series, e.g., the autocorrelation function and the delayed mutual information(DMI)\cite{udaltsov2003cracking,nguimdo2011role}, the filling factor analysis\cite{bunner1998estimation}, extrema statistics\cite{bezruchko2001reconstruction,prokhorov2005reconstruction}, information theory methods\cite{tian1997extraction,azad2002information}, the practical criterion\cite{siefert2007practical} and the permutation entropy and the permutation statistical complexity (${C_{JS}}$)\cite{zunino2010permutation,soriano2011time}.
This work is aim to recover the time delay present in the time-delay chaotic system when the underlying equation is not known and the amount of data is relatively small.
Generally speaking, the phase-space reconstruction is a fundamental tool for chaotic time series analysis. Nonetheless, it is inappropriate to apply this technique to a time-delay system since even a first-order delay differential equation can possess high-dimensional chaotic dynamics\cite{farmer1982chaotic}, and we cannot directly reconstruct the phase-space of such a system since the phase-space of such a system has to be regarded as infinite-dimensional\cite{prokhorov2005reconstruction}. Recently, a new technique, which is called the incomplete reconstruction of the dynamics\cite{garland2015prediction}, gives us new insight into the way to capture the structural information of the dynamics. In the present paper, a new method based on the information captured by the incomplete reconstruction of the dynamics will be introduced to recover the time delay of the system.

Before that, we propose a simple procedure called component permutation, in order to show the local clustering phenomenon of the orbits of the chaotic system in the incomplete two-dimensional reconstructed phase-space.
We find that information captured by the incomplete two-dimensional reconstructed phase-space, not only is related to the time delay present in the system, but also can be transferred to the permutation component by this procedure. Then, in order to recover the time delay of the system, the segmented mean-variance (SMV) is derived from the permutation component.
The proposed SMV will show pronounced maximum when the embedding delay $\tau $ of the incomplete reconstruction is close to the time delay ${\tau _0}$ of the time-delay system.
Numerical data generated from a time-delay system based on the Mackey-Glass system operating in the chaotic regime are used to illustrate the validity of the proposed SMV.
A series of successful time delay identifications demonstrate that the structural information captured by the incomplete two-dimensional reconstructed phase-space is enough to recover the time delay of the system from the scalar time series analysis.
The proposed method is easy to operate with a small amount of computation, and it also has a good robustness against additive observational noise. Most importantly, it can recover the time delay of the system even though the amount of data is relatively small and the feedback strength of the system is weak.

The present paper is structured as follows. In Sec.~\ref{sec2}, the delay-coordinate reconstruction is briefly introduced first, then the component permutation procedure is developed; after that the local clustering phenomenon of the chaotic time series is described by utilizing the numerical data generated by the H\'{e}non map and the Mackey-Glass equation, finally, the segmented mean-variance(SMV) is proposed to recover the time delay present in the scalar time series. In Sec.~\ref{sec3}, the feasibility and reliability of the proposed SMV are first checked by utilizing the numerical time series generated by the time-delay systems based on the the well-known Mackey-Glass equation, then the effects of the additive observational noise, data length and feedback strength on the proposed SMV are tested. At last, a simple comparison of time consumptions between the ${C_{JS}}$ and the proposed SMV for different data lengths is obtained. In Sec.~\ref{sec4}, some brief conclusions are given.
\section{\label{sec2}The local clustering phenomenon and the segmented mean-variance}
\subsection{Delay-coordinate reconstruction}
The reconstruction of the phase-space plays a very important role in chaotic time series analysis since the structure of phase-space is very helpful, and it is also a fundamental tool for nonlinear time series analysis.
The widely used method to reconstruct the phase-space of the dynamics of the chaotic system is the delay-coordinate reconstruction proposed by Takens et al.\cite{takens1981detecting,packard1980geometry}. Specifically, let $\left\{ {{x_n}} \right\}_{n = 1}^N$ be a scalar time series of length $N$, let $m$ be the embedding dimension and $\tau$ be the embedding delay. Then the m-dimensional reconstructed phase-space of this time series consists of the following vectors:
\begin{eqnarray} \label{new}
{{\bf{V}}_n} = \left[ {{x_n},{x_{n + \tau }},{x_{n + 2\tau }}, \cdots {x_{n + \left( {m - 1} \right)\tau }}} \right],
\end{eqnarray}
where $n = 1,2, \cdots ,N - \left( {m - 1} \right)\tau$. Though the embedding dimension $m$ and the embedding delay $\tau$ are very essential for the delay-coordinate reconstruction, good value for them are not easy to be estimated due to the data length, noise, nonstationarity, algorithm parameters and the like\cite{bradley2015nonlinear} in practice. About estimating good value for the embedding dimension $m$ and the embedding delay $\tau$, please see Ref.~\onlinecite{bradley2015nonlinear}, \onlinecite{garland2016leveraging} and the references therein.

Actually, the full structure of the dynamics of the chaotic system is not always necessary, and a partial knowledge of the dynamics is helpful for data analysis purpose sometimes. In Ref.~\onlinecite{garland2015prediction}, it was found that forecast models which utilized the incomplete reconstruction of the dynamics can obtain accurate predictions of the future course of the dynamics. And in Ref.~\onlinecite{Shengli2016detection}, the approach based on the incomplete two-dimensional reconstructed phase-space was successfully applied to distinguish between noise and chaotic signals. In the present paper, we will show that we can recover the time delay ${\tau _0}$ present in the chaotic time series based on the information captured by the incomplete two-dimensional reconstructed phase-space.
\begin{figure}[htbp]
\subfigure[\label{fig1a}]{\includegraphics[width=7.0cm]{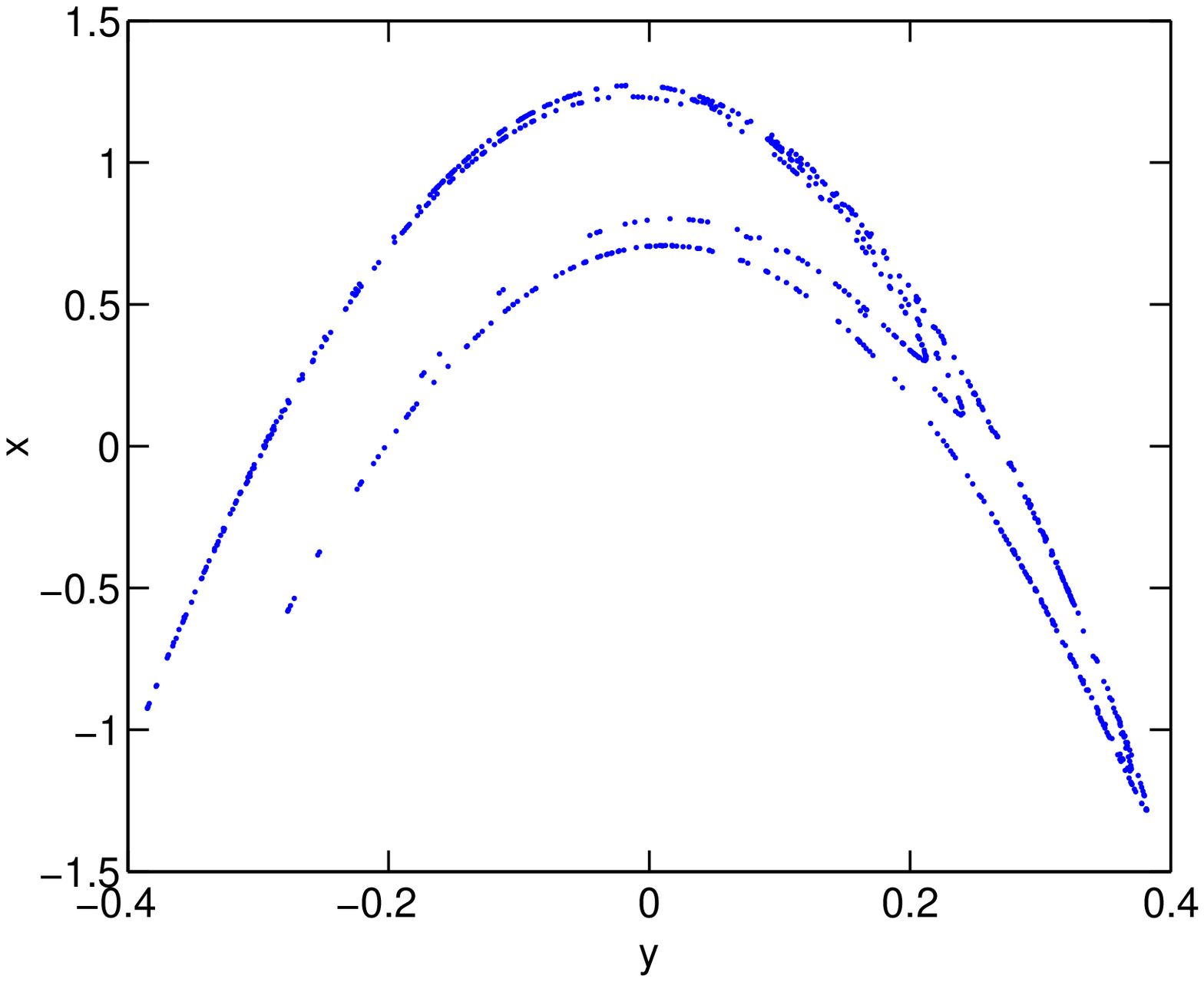}}
\subfigure[\label{fig1b}]{\includegraphics[width=7.0cm]{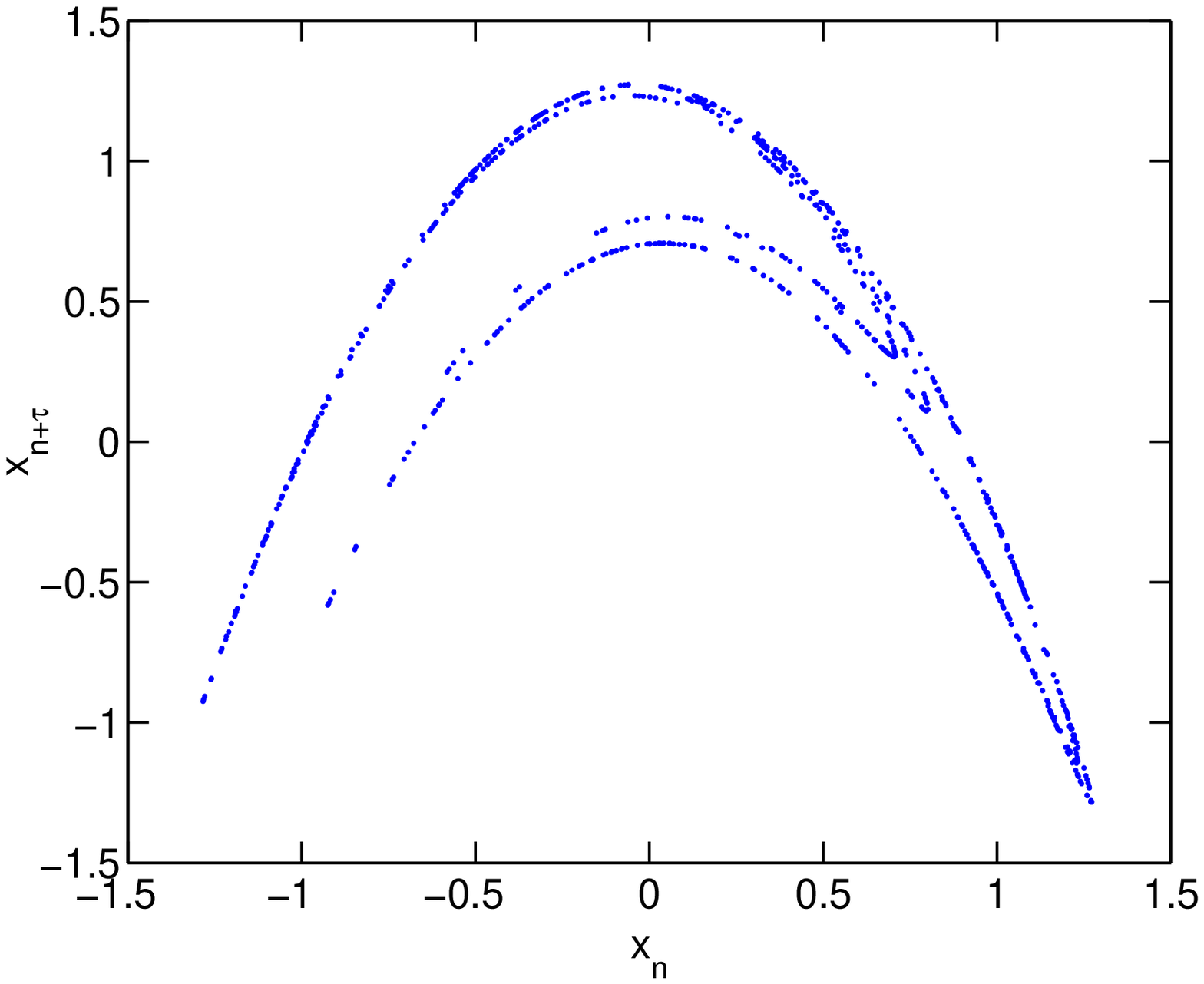}}
\subfigure[\label{fig1c}]{\includegraphics[width=7.0cm]{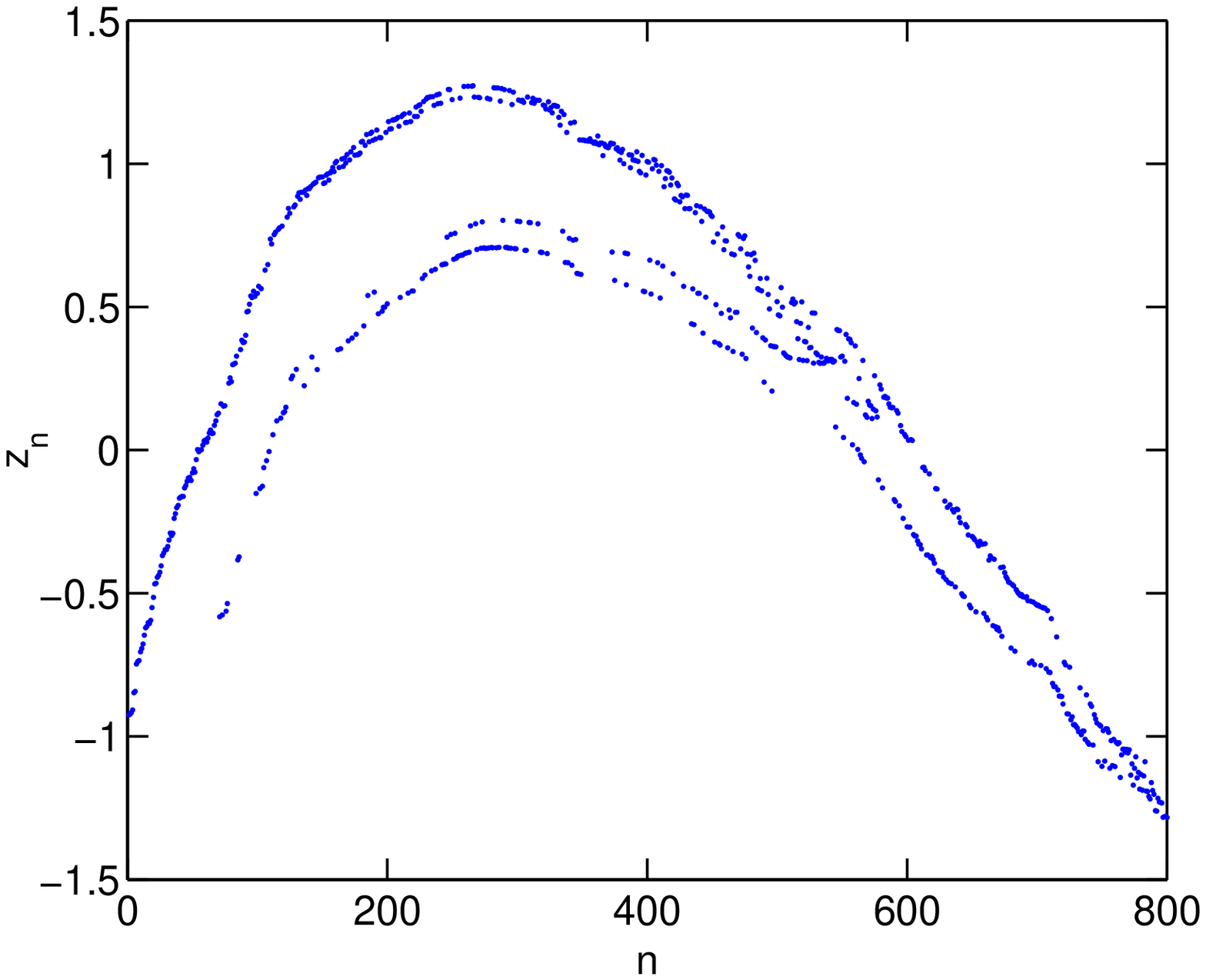}}
\caption{\label{fig1}(a) Phase-space of H\'{e}non map;(b) incomplete two-dimensional reconstructed phase-space of H\'{e}non map(embedding dimension $m = 2$ and embedding delay ${\tau } = 1$);(c) Permutation component with $\tau  = 1$. $N = 800$ data points are used.}
\end{figure}

The incomplete reconstruction means that the choice of embedding dimension $m$ does not satisfy the conditions of Takens' theorem\cite{takens1981detecting}. It should be noted that the full reconstructed phase-space can be obtained for low dimensional chaotic time series if the embedding dimension $m$ and the embedding delay $\tau$ satisfy the conditions of the Takens' theorem. The value of embedding dimension $m$ for the reconstruction of phase-space of the dynamics is set to $2$ in this paper, which means that the phase-space is incompletely reconstructed for high dimensional chaotic systems.
\subsection{The component permutation procedure}
When the embedding dimension $m$ is fixed, another parameter, the embedding delay $\tau$, will have a major impact on the incomplete two-dimensional reconstruction of the phase-space. Obviously, the structure captured by the incomplete two-dimensional reconstructed phase-space is quite diverse if different embedding delay $\tau $ is used. Based on it, we will propose an efficient method to recover the time delay of the system from chaotic time series in the present paper. Before that, a simple but important procedure which is called the component permutation, will be presented first. To the best of our knowledge, this is the first time that the component permutation procedure is introduced.

Let $\left\{ {{x_n}} \right\}_{n = 1}^N$ be a scalar time series of length $N$ and $\tau $ be the embedding delay of the incomplete two-dimensional reconstruction. Then the \textit{component permutation procedure} is implemented as follows:

(1)~Obtaining the first component $\left\{ {x_n^f} \right\}_{n = 1}^{N - \tau } = \left\{ {{x_n}} \right\}_{n = 1}^{N - \tau }$ and the second component $\left\{ {x_n^s} \right\}_{n = 1}^{N - \tau } = \left\{ {{x_n}} \right\}_{n = \tau  + 1}^N$ of the incomplete two-dimensional reconstructed phase-space,

(2)~Sorting the first component $\left\{ {x_n^f} \right\}_{n = 1}^{N - \tau }$ with ascending order, let ${n_{new}}$ be the new subscript after sorting. Let ${z_n} = x_{{n_{new}}}^s$, then $\left\{ {{z_n}} \right\}_{n = 1}^{N - \tau }$ will be a new time series.

We will call $\left\{ {{z_n}} \right\}_{n = 1}^{N - \tau }$ the \textit{permutation component} for the reason that it is actually a permutation of the second component $\left\{ {x_n^s} \right\}_{n = 1}^{N - \tau }$. The \textit{component permutation procedure} can be better described with a simple example; suppose that we have a short time series $\left\{ {{x_n}} \right\}_{n = 1}^7 = \left\{ {1.1,7.1,6.1,2.3,4.5,5.3,8.2} \right\}$ and we set the embedding delay $\tau = 1$, then we can obtain the first component $\left\{ {x_n^f} \right\}_{n = 1}^6 = \left\{ {1.1,7.1,6.1,2.3,4.5,5.3} \right\}$ and the second component $\left\{ {x_n^s} \right\}_{n = 1}^6 = \left\{ {7.1,6.1,2.3,4.5,5.3,8.2} \right\}$ of the original time series. By sorting the first component with ascending order, we can get the new subscript $n_{new} =1,4,5,6,3,2$; correspondingly, we can obtain the permutation component $\left\{ {{z_n}} \right\}_{n = 1}^6 = \left\{ {7.1,4.5,5.3,8.2,2.3,6.1} \right\}$ by utilising the new subscript.
\begin{figure*}[htbp]
\centering
\subfigure[\label{fig2a}]{\includegraphics[width=7.0cm]{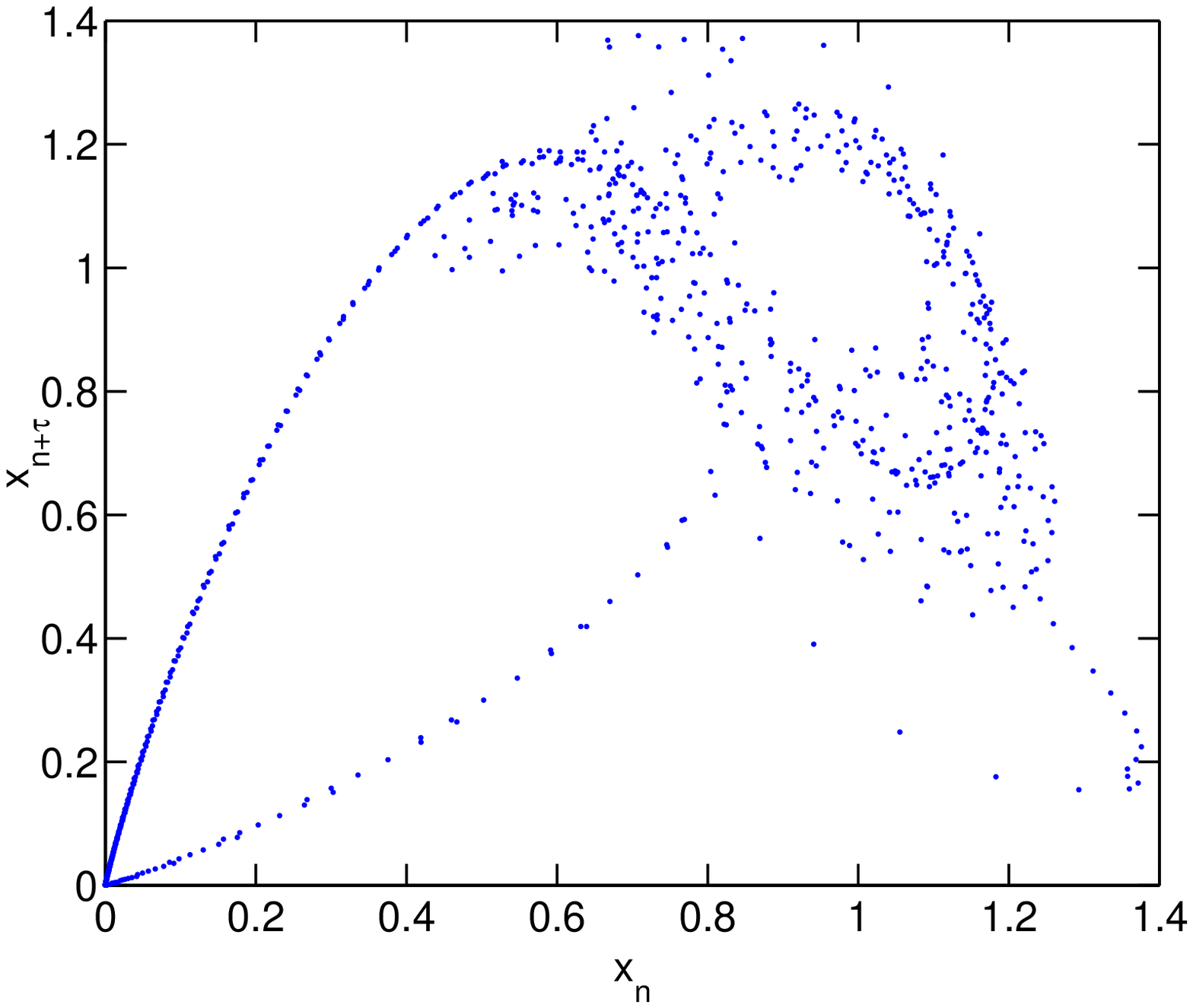}}
\subfigure[\label{fig2b}]{\includegraphics[width=7.0cm]{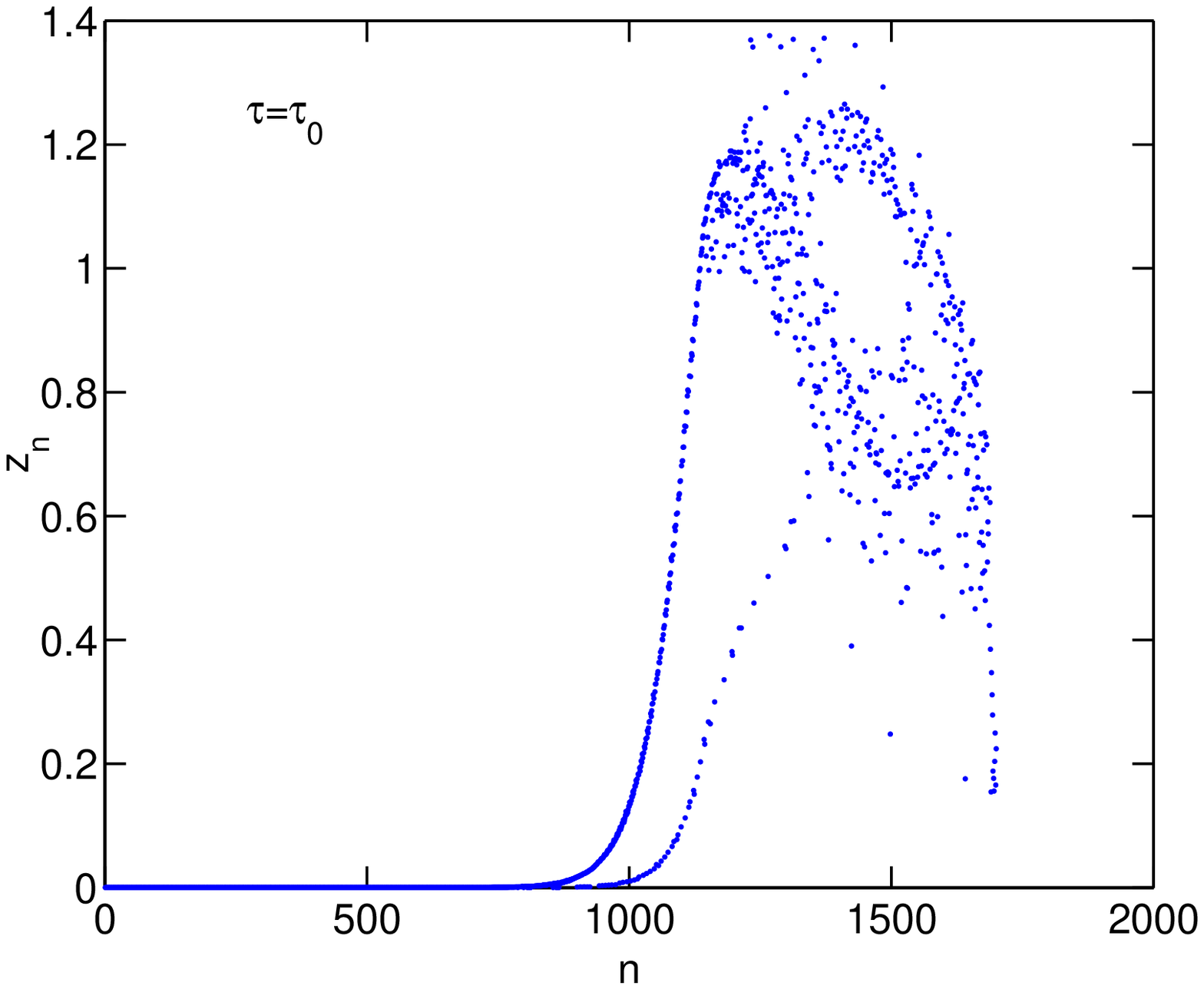}}
\subfigure[\label{fig2c}]{\includegraphics[width=7.0cm]{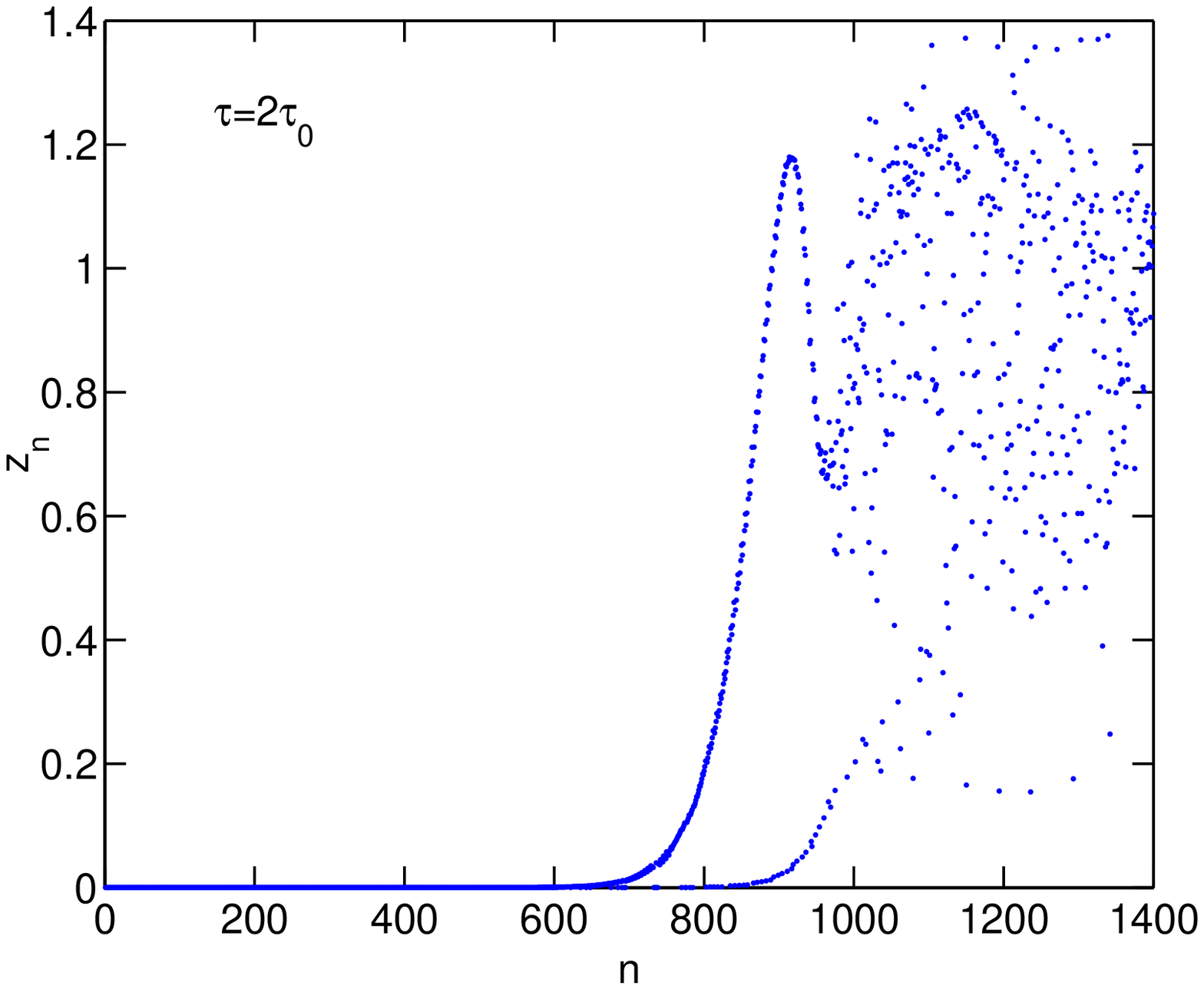}}
\subfigure[\label{fig2d}]{\includegraphics[width=7.0cm]{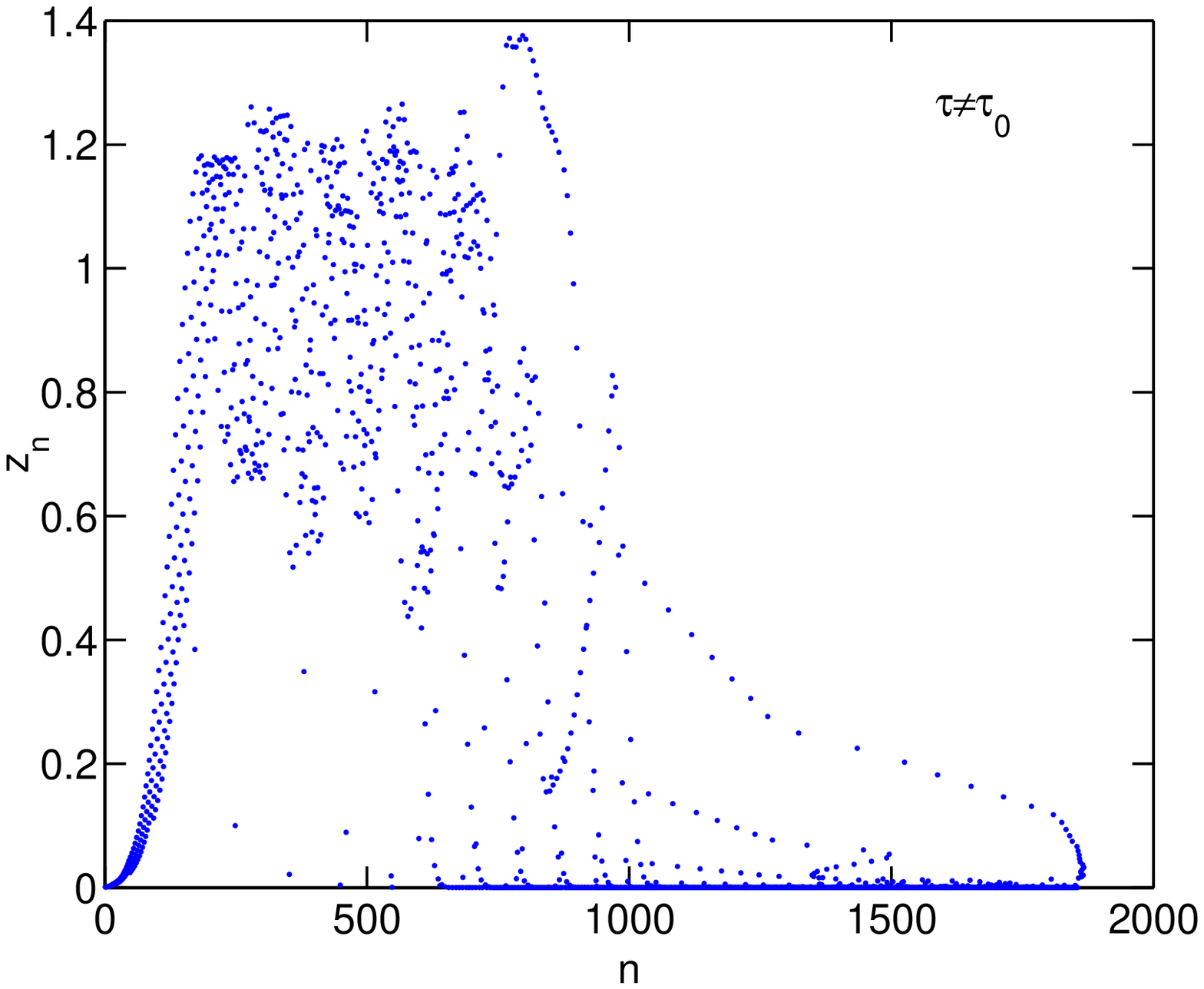}}
\caption{\label{fig2} (a) Incomplete two-dimensional reconstructed phase-space of Mackey-Glass time series (embedding dimension $m = 2$ and embedding delay $\tau  = 300$) and permutation component with embedding delay (b) $\tau  = 300$ (c) $\tau  = 600$ (d) $\tau  = 133$.}
\end{figure*}
\subsection{The local clustering phenomenon}
After obtaining the permutation component, we have observed that there is a special relationship between the incomplete two-dimensional reconstructed phase-space and its corresponding permutation component. To better describe this relationship, we first consider the following equation for H\'{e}non map\cite{henon1976two}:
\begin{eqnarray} \label{eqone}
\left\{ \begin{array}{lc}
{x_{n + 1}} = {y_n} + 1 - ax_n^2\\
{y_{n+1}} = b{x_n}
\end{array} \right.
\end{eqnarray}

The typical values $a = 1.4$ and $b = 0.3$ are chosen to produce a deterministic chaotic time series. The H\'{e}non map is used here because that it has a low-dimensional attractor and the relationship can be described more clearly.
It should be stressed that the discrete chaotic map itself is not the focus of this paper.

In Fig.~\ref{fig1} we plot the phase-space, the incomplete two-dimensional reconstructed phase-space and the corresponding permutation component of H\'{e}non map. From Fig.~\ref{fig1}(a) and Fig.~\ref{fig1}(b) it can be observed that the incomplete reconstruction recovers all of the structure of this system since the delay-coordinate reconstruction of this data with $m=2$ and $\tau = 1$ is indeed the actual map just scaled. However, the main point of Fig.~\ref{fig1}, concerns the similarity between Fig.~\ref{fig1}(b) and Fig.~\ref{fig1}(c), which represent the incomplete two-dimensional reconstructed phase-space and its corresponding permutation component, respectively. By studying the permutation component, we also find that the values of the adjacent data of the permutation component are very close when the similarity is well displayed, as if these data are locally clustered by the component permutation procedure.
We will call that the \textit{local clustering phenomenon} in this paper. In other words, the local clustering phenomenon contains two meanings, they are, the similarity between the incomplete two-dimensional reconstructed phase-space and its corresponding permutation component, and the clustering of the data in the permutation component.

The similarity between the incomplete two-dimensional reconstructed phase-space and the corresponding permutation component of H\'{e}non map is perfectly displayed since the phase-space of H\'{e}non map is well reconstructed. It should be noted that the similarity exists when other values of $\tau$ are used and disappears when $\tau $ is large enough for H\'{e}non map, which we do not plot in this paper. It seems that the local clustering phenomenon depends on the embedding delay $\tau$ of the incomplete reconstruction.
However, for a time-delay chaotic system which can produce highly complex dynamics, the incomplete two-dimensional reconstructed phase-space can not capture the full structure of the dynamics, does the local clustering phenomenon still exist? To answer this question, we consider the well-known Mackey-Glass equation\cite{mackey1977oscillation}:
\begin{eqnarray}
\begin{array}{*{20}{c}}
{\frac{{dx}}{{dt}} = \frac{{\alpha x\left( {t - {\tau _0}} \right)}}{{1 + {x^\gamma }\left( {t - {\tau _0}} \right)}} - x,}&{\alpha ,\gamma  > 0,}\label{equationmg}
\end{array}
\end{eqnarray}
where ${\tau _0}$ is the time delay feedback, $\alpha $ is the feedback strength, $\gamma $ is the degree of nonlinearity and $t$ is a dimensionless time. The typical values ${\tau _0}{\rm{ = }}60,\alpha  = 2$ and $\gamma  = 10$ are chosen to make the system operate in the chaotic regime. For the purpose of obtaining the Mackey-Glass time series, the forth-order Runge\textrm{-}Kutta method\cite{gard1988introduction} is used to numerically integrate the equation from Eq.~(\ref{equationmg}), and the integration step and sampling step are $\Delta t = 0.{\rm{0}}1$ and $\delta t = 0.2$ time units, respectively. The time delay present in the Mackey-Glass time series is 300 $\left({{{\tau _0}} \mathord{\left/ {\vphantom {{{\tau _0}} {\delta t}}} \right. \kern-\nulldelimiterspace} {\delta t}} = 300\right)$ under these parameters.

In Fig.~\ref{fig2} we plot the incomplete two-dimensional reconstructed phase-space with embedding delay $\tau=300$ and its corresponding permutation component of the Mackey-Glass time series. Besides, we also plot the permutation component with embedding delay $\tau  = 600$ and $\tau  = 133$ in Fig.~\ref{fig2}, while their corresponding incomplete reconstructed phase-space are not plotted for the sake of brevity. From Fig.~\ref{fig2}(a) and Fig.~\ref{fig2}(b) it can be observed that the local clustering phenomenon still exists, though it is poorly displayed (compared with the H\'{e}non scenario).

The local clustering phenomenon illustrates that the information captured by the incomplete two-dimensional reconstructed phase-space is transferred to the corresponding permutation component by the procedure of component permutation. Hence, we can extract some important features of the dynamics of the time-delay chaotic system from the corresponding permutation component. It is also more convenient to study the permutation component rather than the incomplete reconstructed phase-space since the dimension needs to be tackled is reduced.

As shown in Fig.~\ref{fig2}, the permutation component has three different forms when different embedding delays $\tau$ of the incomplete reconstruction are considered, and these forms are related to the time delay ${\tau_0}$ of the system. The relationships between the embedding delay $\tau$ of the incomplete reconstruction and the time delay ${\tau_0}$ present in the system are described as below.
\begin{eqnarray}
\begin{array}{l}
{\rm{I}}\;\tau  = {\tau _0}\\
{\rm{II}}\;\tau  = n{\tau _0},n = 2,3, \cdots \\
{\rm{III}}\;\tau  \ne n{\tau _0},n = 1,2,3, \cdots
\end{array}
\label{relationship}
\end{eqnarray}
In case I, i.e., the embedding delay $\tau$ of the incomplete reconstruction is equal to the time delay ${\tau _0}$ of the system, the amount of information captured by the incomplete reconstructed phase-space is the most, as shown in Fig.~\ref{fig2}(a). Accordingly, the information contained in the permutation component is the most and the local clustering phenomenon is well displayed, as shown in Fig.~\ref{fig2}(b).
In case II, the permutation component also shows some fundamental structure of the dynamics of the system since the structural information captured by the incomplete two-dimensional reconstructed phase-space is also considerable, as shown in Fig.~\ref{fig2}(c).
In case III, however, the permutation component acts like random time series for the reason that the incomplete two-dimensional reconstructed phase-space can not capture any structural information of the dynamics of the system, as shown in Fig.~\ref{fig2}(d). It should be noted that there is no local clustering phenomenon at this time.
From Fig.~\ref{fig2} and Fig.~\ref{fig1} we also observe that the local clustering phenomenon of the Mackey-Glass time series is not demonstrated as well as that of the H\'{e}non map, the reason is that the incomplete two-dimensional reconstructed phase-space can not capture the full dynamics of the Mackey-Glass system.
\subsection{The segmented mean-variance}
As described above, for a time-delay chaotic system, the property of the permutation component is closely related to the relationship between the embedding delay $\tau$ of the incomplete reconstruction and the time delay ${\tau _0}$ of the system. We find that the local clustering phenomenon is well displayed when the embedding delay $\tau $ of the incomplete reconstruction is equal to the time delay ${\tau _0}$ present in the system since the structural information captured by the incomplete two-dimensional reconstructed phase-space is the most in this situation.
Inspired by the local clustering phenomenon, we present a novel and simple approach to recover the time delay $\tau $ of the system from the permutation component in this paper. This method, which will be called the segmented mean-variance (SMV), is based on the calculation of the mean and variance of the permutation component. The calculation procedure of the SMV is described as follows:

(1)~Dividing the permutation component $\left\{ {{z_n}} \right\}_{n = 1}^{N - \tau }$ into $L$ groups: ${Z_l} = \left\{ {{z_{\left( {l - 1} \right)K + 1}},{z_{\left( {l - 1} \right)K + 2}}, \cdots ,{z_{lK}}} \right\},l = 1,2, \cdots ,L$ , with $K = \left\lfloor {{N \mathord{\left/
 {\vphantom {N L}} \right.
 \kern-\nulldelimiterspace} L}} \right\rfloor $ being the amount of data of each group and $\left\lfloor P \right\rfloor$ denoting an integer less than or equal to $P$;

(2)~Calculating the mean ${\hat \mu _l}$ and the variance $\hat \sigma _l^2$ of each group ${Z_l},l = 1,2, \cdots ,L$,
 \begin{eqnarray}
 \begin{array}{l}
{{\hat \mu }_l} = \frac{1}{K}\sum\limits_{k = 1}^K {{z_{\left( {l - 1} \right)K + k}}} ,\\
\hat \sigma _l^2 = \frac{1}{{K - 1}}\sum\limits_{k = 1}^K {{{\left( {{z_{\left( {l - 1} \right)K + k}} - {{\hat \mu }_l}} \right)}^2}} ;
\end{array}
\end{eqnarray}

(3)~Computing the mean of ${\hat \mu _l}$ and $\hat \sigma _l^2,l = 1,2, \cdots ,L$,
\begin{eqnarray}
\hat \mu  = \frac{1}{L}\sum\limits_{l = 1}^L {{{\hat \mu }_l}} ,{\hat \sigma ^2} = \frac{1}{L}\sum\limits_{l = 1}^L {\hat \sigma _l^2} ;
\end{eqnarray}

(4)~Calculating the variance of ${\hat \mu _l},l = 1,2, \cdots ,L$,
\begin{eqnarray}
\hat \sigma _0^2 = \frac{1}{{L - 1}}\sum\limits_{l = 1}^L {{{\left( {{{\hat \mu }_l} - \hat \mu } \right)}^2}} ;
\label{equationsigm0}
\end{eqnarray}

(5)~Then the SMV is obtained as below,
\begin{eqnarray}
SMV = \frac{K}{{{{\hat \sigma }^2}}}\hat \sigma _0^2.
\label{equationka}
\end{eqnarray}
According to the relationships described in Eq.~\ref{relationship}, the SMV will show three different types of values.
In the first case, the values of data of each group ${Z_l},l = 1,2, \cdots ,L$ are almost the same because of the well presented local clustering phenomenon, therefore, the mean of segmented variances ${\hat \sigma ^2}$ is of small value. Meanwhile, the values of data of different groups are different, then the segmented means $\hat \sigma _0^2$ is of large value. A small value of ${\hat \sigma ^2}$ and a large value of  $\hat \sigma _0^2$ will lead to a large SMV in this situation. It should be noted that the parameter $K$ in Eq.~(\ref{equationka}) is just a multiplication factor which makes the SMV follow F distribution if the time series $\left\{ {{x_n}} \right\}_{n = 1}^N$ is a Gaussian white noise\cite{Shengli2016detection}.

In the second case, though the local clustering phenomenon is not presented as well as that of the first case, the incomplete two-dimensional reconstructed phase-space also can capture some fundamental information of the dynamics of the system. Though the value of the SMV is large enough, but it still much smaller than that of the first case.
In the last case, there is no helpful information contained in the permutation component, and the local clustering phenomenon disappears. The permutation component acts like a random time series. Thus, the mean of segmented variances ${\hat \sigma ^2}$ is of large value and the variance of segmented means $\hat \sigma _0^2$ is of small value, so the value of the proposed SMV is relatively small. Above all, we can recover the time delay present in the system according to the values of the SMV.
\section{\label{sec3}NUMERICAL RESULTS AND DISCUSSIONS}
As stated in Sec.~\ref{sec2}, the values of the proposed SMV are relatively different according to the relationship between the embedding delay $\tau $ of the incomplete reconstruction and the time delay ${\tau _0}$ of the system, which means that we can identify the time delay present in the scalar time series according to the values of the proposed SMV. In this section, some experiments will be given in order to check the effectiveness and reliability of the proposed SMV. Besides, for the purpose of comparison, the permutation statistical complexity (${C_{JS}}$)\cite{zunino2010permutation} is used as the gold standard.
The ${C_{JS}}$ is a relatively new method proposed to identify the time delay of the system and its performance is relatively good\cite{zunino2010permutation}. Note that the calculation of ${C_{JS}}$ is related to the phase-space reconstruction\cite{rosso2007distinguishing}, so it also need two parameters, namely the embedding delay $\tau$ and embedding dimension $m$ when evaluating the ${C_{JS}}$.
Generally speaking, to obtain a reliable statistics when evaluating the ${C_{JS}}$, the length $N$ of the time series and the embedding dimension $m$ should satisfy the condition $N \gg m!$\cite{staniek2007parameter}, or they should satisfy the condition $N \ge 5m!$\cite{matilla2008non} at least. In the following experiments, the embedding dimensions for ${C_{JS}}$ are chosen according to the latter condition. About this approach, please see Ref.~\onlinecite{rosso2007distinguishing} and \onlinecite{zunino2010permutation} for details. It should be pointed out that the starting points of the ${C_{JS}}$ and proposed SMV are totally different though they  both are related to the phase-space reconstruction.
\begin{figure}[htbp!]
\centering
\subfigure[\label{fig3a}]{\includegraphics[width=7.0cm]{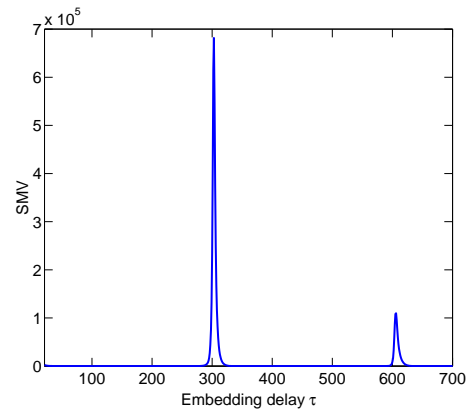}}
\subfigure[\label{fig3b}]{\includegraphics[width=7.0cm]{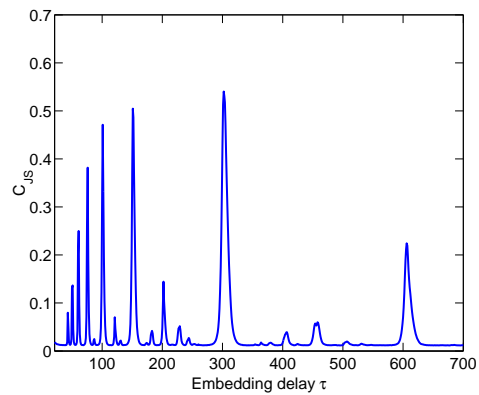}}
\caption{\label{fig3} (a) SMV as a function of the embedding delay $\tau $ for number of data segments $L = 5$. (b) the ${C_{JS}}$ as a function of the embedding delay $\tau $ for embedding dimension $m = 8$. $N = {10^6}$ data points are used.}
\end{figure}
\begin{figure}[htbp!]
\subfigure[\label{fig4a}]{\includegraphics[width=7.0cm]{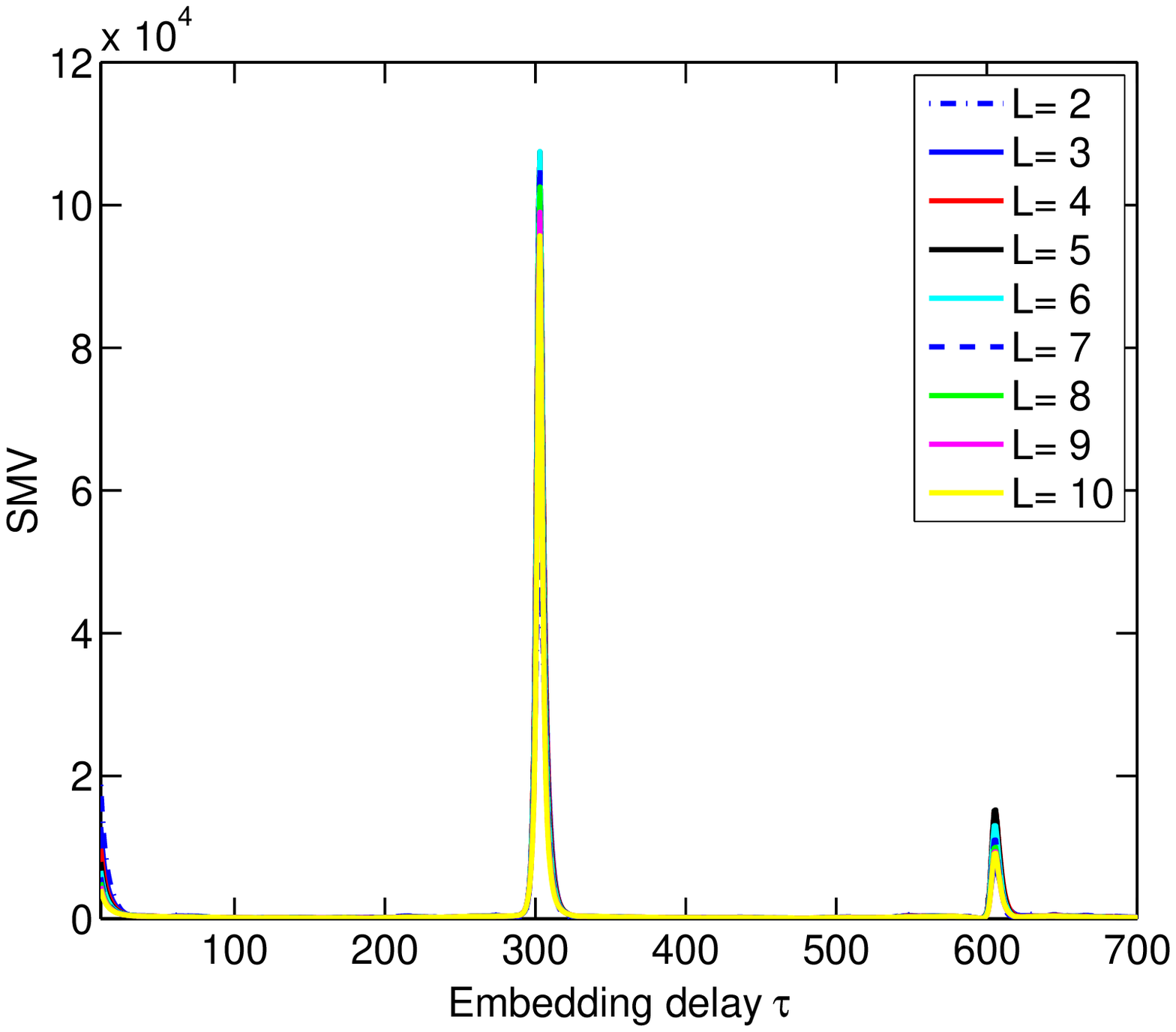}}
\subfigure[\label{fig4b}]{\includegraphics[width=7.0cm]{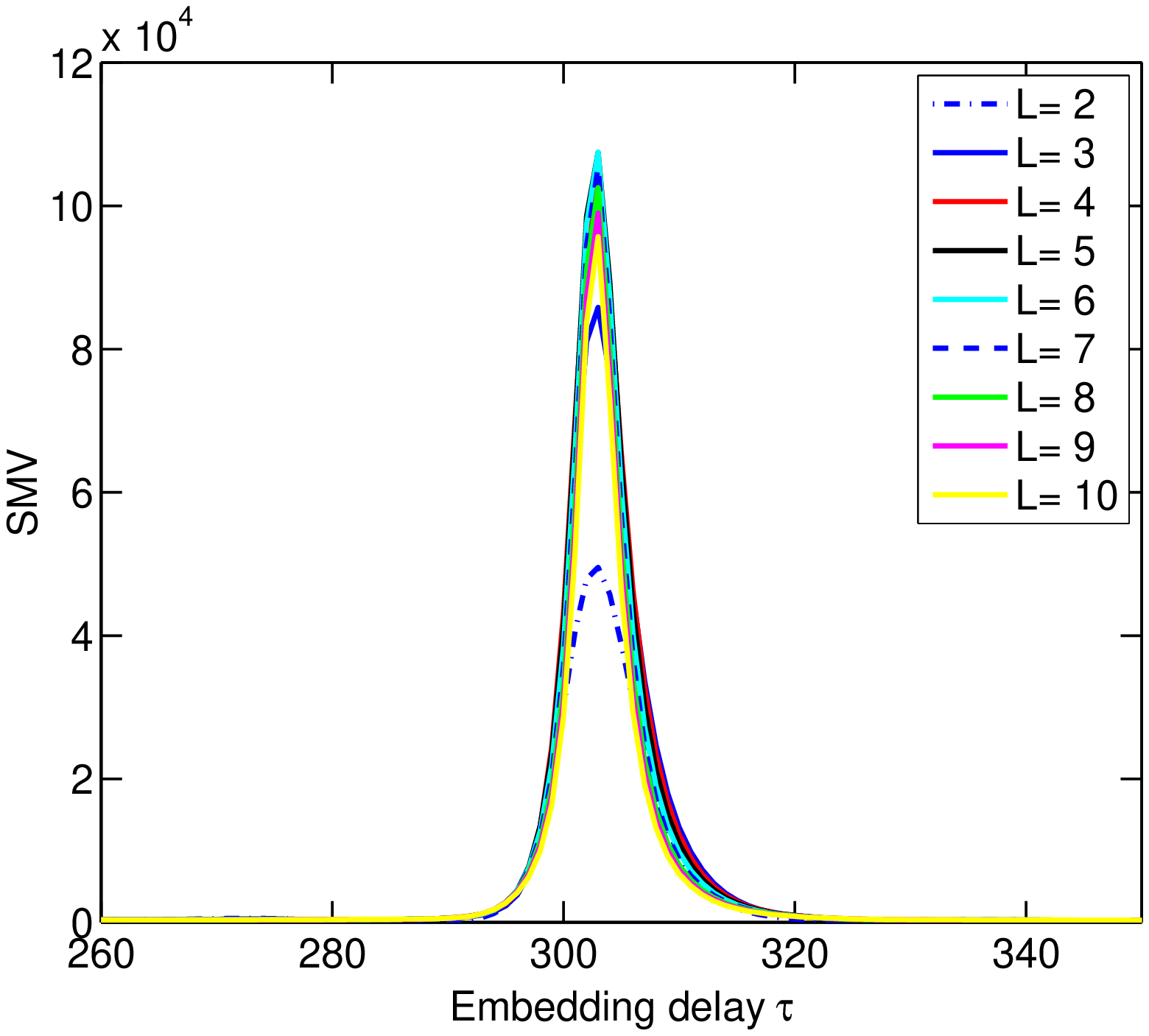}}
\caption{\label{fig4} (a) SMV as a function of embedding delay $\tau $ for the numbers of data segments $2 \le L \le 10$. (b) Enlargement near the time delay ${\tau _0}$ of the system in order to observe more clearly the effect of the number of data segments $L$ on the SMV. $N = 2 \times {10^5}$ data points are used.}
\end{figure}
\begin{figure}[htbp!]
\includegraphics[width=7.0cm]{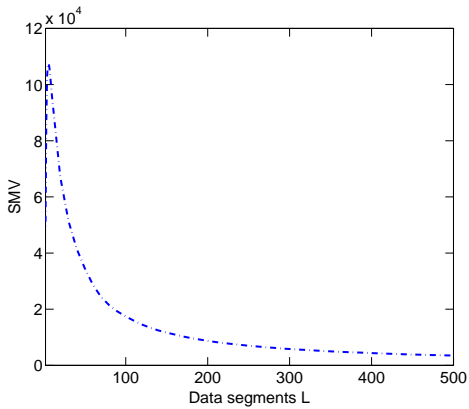}
\caption{\label{fig5} SMV as a function of number of data segments $ L $ for embedding delay $\tau = 303$. The maximum value of the SMV occurs when $L = 6$. $N = 2 \times {10^5}$ data points.}
\end{figure}

Numerical data used in the following simulations are generated by the time-delay systems based on the Mackey-Glass equation.
First of all, the effectiveness of the SMV will be tested by calculating the SMV as a function of the embedding delay $\tau $ of the incomplete reconstruction, and the effect of the number of data segments $L$ on the proposed SMV will be discussed. Then, the effect of additive observational noise, data length and feedback strength on the proposed SMV will be checked. Finally, the time complexity of the proposed SMV is obtained for different data lengths.
\subsection{The effectiveness of the SMV}
In order to check the effectiveness of the proposed method, we calculate the SMV as a function of the embedding delay $\tau $ of the incomplete reconstruction for the Mackey-Glass time series generated by Eq.~(\ref{equationmg}).
The results are shown in Fig.~\ref{fig3}(a). It can be clearly observed that the proposed SMV has well-defined and sharp maxima when the embedding delay $\tau $ of the incomplete reconstruction is close to the time delay ${\tau _0}$ of the system, i.e. for $\tau $ near 300$\left({{{\tau _0}} \mathord{\left/ {\vphantom {{{\tau _0}} {\delta t}}} \right. \kern-\nulldelimiterspace} {\delta t}} = 300\right)$. In this situation, the structural information of the system captured by the incomplete two-dimensional reconstructed phase-space is the most and the local clustering phenomenon is well presented. Consequently, the value of the SMV is the largest. Moreover, the proposed SMV also shows clear peak when the embedding delay $\tau $ of the incomplete reconstruction is approximately double the time delay ${\tau _0}$ of the system, i.e. for $\tau $ near 600$\left({{{\tau _0}} \mathord{\left/ {\vphantom {{{\tau _0}} {\delta t}}} \right. \kern-\nulldelimiterspace} {\delta t}} = 600\right)$. While, the value of the proposed SMV is much smaller in the case of other embedding delays. Thus, the results shown in Fig.~\ref{fig3}(a) are perfectly consistent with the discussion in Sec.~\ref{sec2}.

From Fig.~\ref{fig3}(a) it can also be observed that there is a light time delay overestimation. This time delay overestimation can be attributed to the internal response time or inertia of the Mackey-Glass system\cite{zunino2010permutation}. It is difficult to estimate the inertia accurately, and most of the methods proposed to recover the time delay present in the recorded time series\cite{rontani2007loss,rontani2009time}, are affected by the inertia. We also calculate the ${C_{JS}}$ as a function of the embedding delay $\tau$ for the reason that the ${C_{JS}}$ is also affected by the inertia and for comparison purpose, the results are shown in Fig.~\ref{fig3}(b). It can be observed from Fig.~\ref{fig3}(b) that there are a lot of spurious peaks in the ${C_{JS}}$, while these spurious peaks do not appear in our approach.

The number of data segments $L$ is important for calculation of the proposed SMV. It should be pointed out that $L$ should satisfy the following condition
\begin{eqnarray}
 2 \le L \le \left\lfloor {{{\left( {N - {\tau _{\max }}} \right)} \mathord{\left/
 {\vphantom {{\left( {N - {\tau _{\max }}} \right)} 2}} \right.
 \kern-\nulldelimiterspace} 2}} \right\rfloor,
 \label{condition_L}
\end{eqnarray}
 where ${\tau _{\max }}$ is the largest embedding delay of the incomplete reconstruction one chooses and $N$ is the data length. From Eq.~(\ref{equationsigm0}) we can find that $L \ne 1$ is obvious, and if $L > \left\lfloor {{{\left( {N - {\tau _{\max }}} \right)} \mathord{\left/
 {\vphantom {{\left( {N - {\tau _{\max }}} \right)} 2}} \right.
 \kern-\nulldelimiterspace} 2}} \right\rfloor $, the amount of data in each group is $K=1$, the calculation of SMV is meaningless in this situation. It is obvious that the values of the proposed SMV are different if we choose different numbers of data segments $L$. In order to check the effect of $L$ on the SMV, in Fig.~\ref{fig4}(a) we plot the SMV as a function of embedding delay $\tau $ of the incomplete reconstruction for different numbers of data segments $L$, and and in Fig.~\ref{fig4}(b) we plot the
enlargement near the time delay ${{\tau _0}}$ the system in order to observe more clearly. From Fig.~\ref{fig4}, it can be observed that the values of the SMV corresponding to $L = 5$ and $L = 6$ are bigger than that of other data segments $L$. To better explain this, the proposed SMV as a function of the number of data segments $L$ for $\tau = 303$ (because of the time delay overestimation) is plotted in Fig.~\ref{fig5}, it can be observed that the SMV is maximized when $L = 6$ in this situation (The time series is generated from Eq.~(\ref{equationmg}) and $N = 2 \times {10^5}$ data are used.), while the corresponding number of data segments may not be $L = 6$ for other data lengths and time series derived from other time-delay chaotic systems. Nevertheless, the proposed SMV can always recover the time delay of the system as long as the number of data segments $L$ one chooses satisfies the condition in Eq.~(\ref{condition_L}). From now on, the number of data segments $L$ will be fixed to 5 for the sake of uniformity in this article.
\begin{figure}[hbtb!]
\centering
\subfigure[\label{fig6a}]{\includegraphics[width=7.0cm]{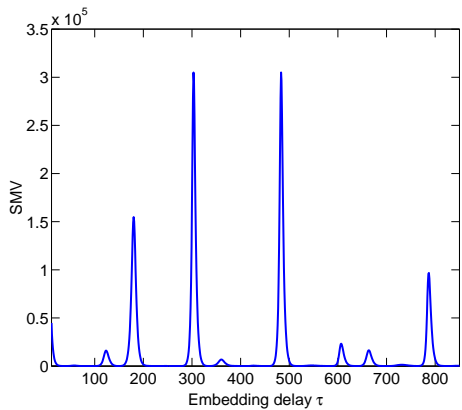}}
\subfigure[\label{fig6b}]{\includegraphics[width=7.0cm]{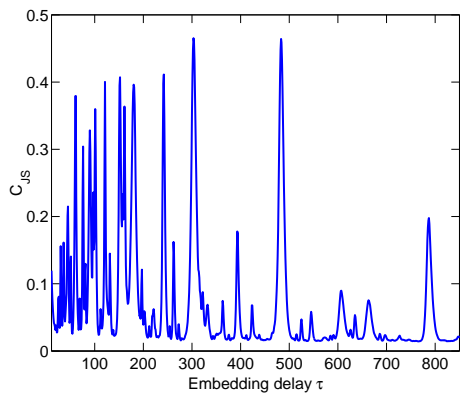}}
\caption{\label{fig6}(a) SMV and (b) the ${C_{JS}}$ as a function of the embedding delay $\tau $ for a Mackey-Glass system with time delays ${\tau _{0,1}} = 60$ and ${\tau _{0,2}} = 96$. The number of data segments for the SMV is $L = 5 $, the embedding dimension for ${C_{JS}}$ is $m = 8$, $N = 1 \times {10^6}$ data points are used.}
\end{figure}

In practical applications, there are more than one time delay in the system sometimes. In order to test the performance of the SMV in this case, we consider the generalized Mackey-Glass equation\cite{voss1997reconstruction} with two time delays:
\begin{eqnarray}
\frac{{dx}}{{dt}} = \frac{1}{2}\sum\limits_{k = 1}^2 {\frac{{\alpha x\left( {t - {\tau _{0,k}}} \right)}}{{1 + {x^\gamma }\left( {t - {\tau _{0,k}}} \right)}}}  - x,
\end{eqnarray}
To obtain the numerical data, the same integration method and parameters $\left(\alpha  = 2,\gamma  = 10\right)$ as in the single time delay case are used.
\begin{figure}[htbp!]
\subfigure[\label{fig7a}]{\includegraphics[width=7.0cm]{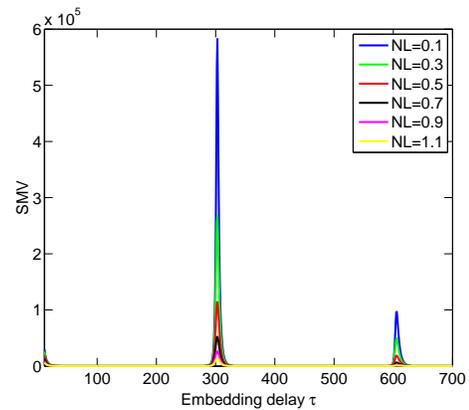}}
\subfigure[\label{fig7b}]{\includegraphics[width=7.0cm]{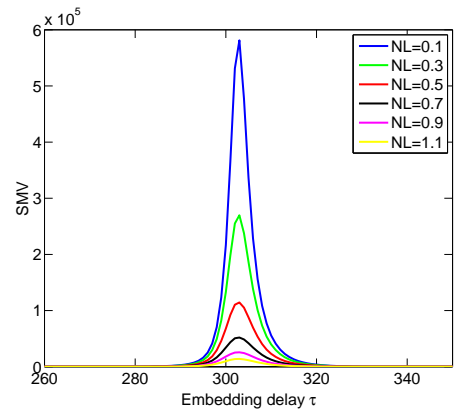}}
\caption{\label{fig7} (a) SMV as a function of the embedding delay $\tau $ of the incomplete reconstruction for different levels of the observational noise. The noise levels associated with the different curves ($NL = 0.1,0.3,0.5,0.7,0.9,1.1$) increases from top to bottom. The number of data segments is $L = 5$  and $N = {10^{\rm{6}}}$ data points are used. (b) Enlargement near the time delay ${\tau_0}$ of the system in order to observe more clearly the effect of the $NL$ on the SMV.}
\end{figure}
In Fig.~\ref{fig6}(a) we plot the SMV as a function of the embedding delay $\tau $ of the incomplete reconstruction in the case of a generalized Mackey-Glass system with two time delays$\left({\tau _{0,1}} = 60,{\tau _{0,2}} = 96\right)$. The proposed SMV shows obvious peaks when the embedding delays of the incomplete reconstruction are close to the time delays of the system, i.e. $\tau  \sim 300$ $\left({{{\tau _{0,1}}} \mathord{\left/{\vphantom {{{\tau _{0,1}}} {\delta t}}} \right.\kern-\nulldelimiterspace} {\delta t}} = 300\right)$ and $\tau  \sim 480$ $\left({{{\tau _{0,2}}} \mathord{\left/{\vphantom {{{\tau _{0,2}}} {\delta t}}} \right. \kern-\nulldelimiterspace} {\delta t}} = 480\right)$. Similar to the case of one time delay, there is also a slight time
delay overestimation.
From Fig.~\ref{fig6}(a) it can also be seen that the SMV shows peaks when the embedding delay is close to $\tau  = 180$($480-300=180$) and $\tau  = 780$($480+300=780$), but less pronounced. The occurrence of these peaks is similar to the intermodulation in inverters; their presence do not affect the estimation of the time delays of the system. As a comparison, the ${C_{JS}}$ is also calculated from the same time series, the result is shown in Fig.~\ref{fig6}(b). From Fig.~\ref{fig6}(b) we can also observe the intermodulation phenomena, besides, a lot of spurious peaks appeared on the left side of the time delays of the system, which has a bad effect on the identification of the time delays of the system. Thus, it can be concluded that our method still works well in the case of two time delays, and its performance is better than that of the ${C_{JS}}$.
\subsection{The effect of an additive observational noise}
The next goal of this paper is to analyze the effect of an additive observational noise on the proposed method. It is meaningful since the experimental time series is always contaminated by observational noise in practice. For this purpose, a Gaussian white noise with different noise levels is added to the numerical data generated by the Mackey-Glass system with one time delay. The noise level (NL) is defined as the ratio of the standard deviation of the noise and the standard deviation of the original signal.

\begin{figure*}[htbp!]
\centering
\subfigure[\label{fig8a}]{\includegraphics[width=7.0cm]{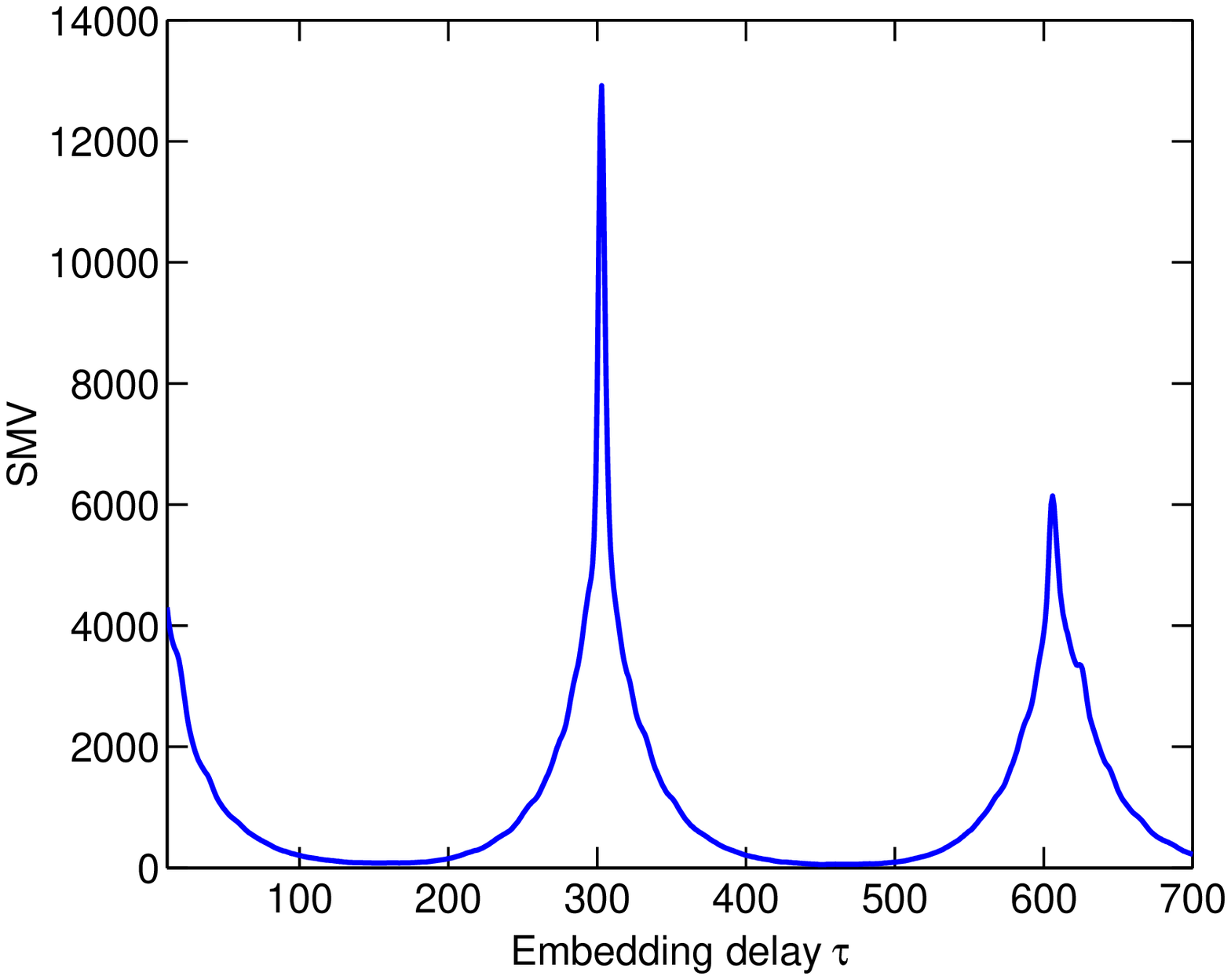}}
\subfigure[\label{fig8b}]{\includegraphics[width=7.0cm]{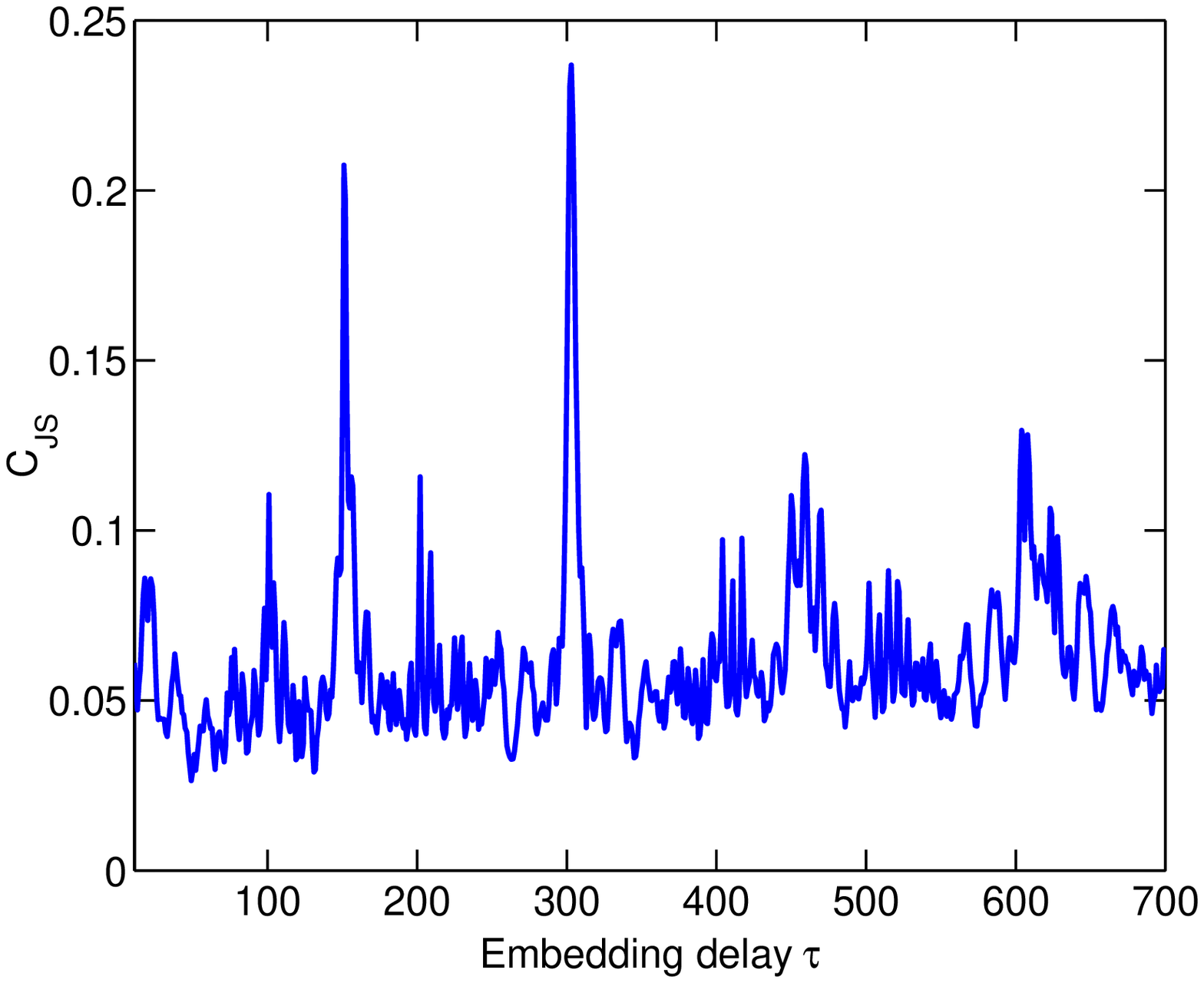}}
\subfigure[\label{fig8c}]{\includegraphics[width=7.0cm]{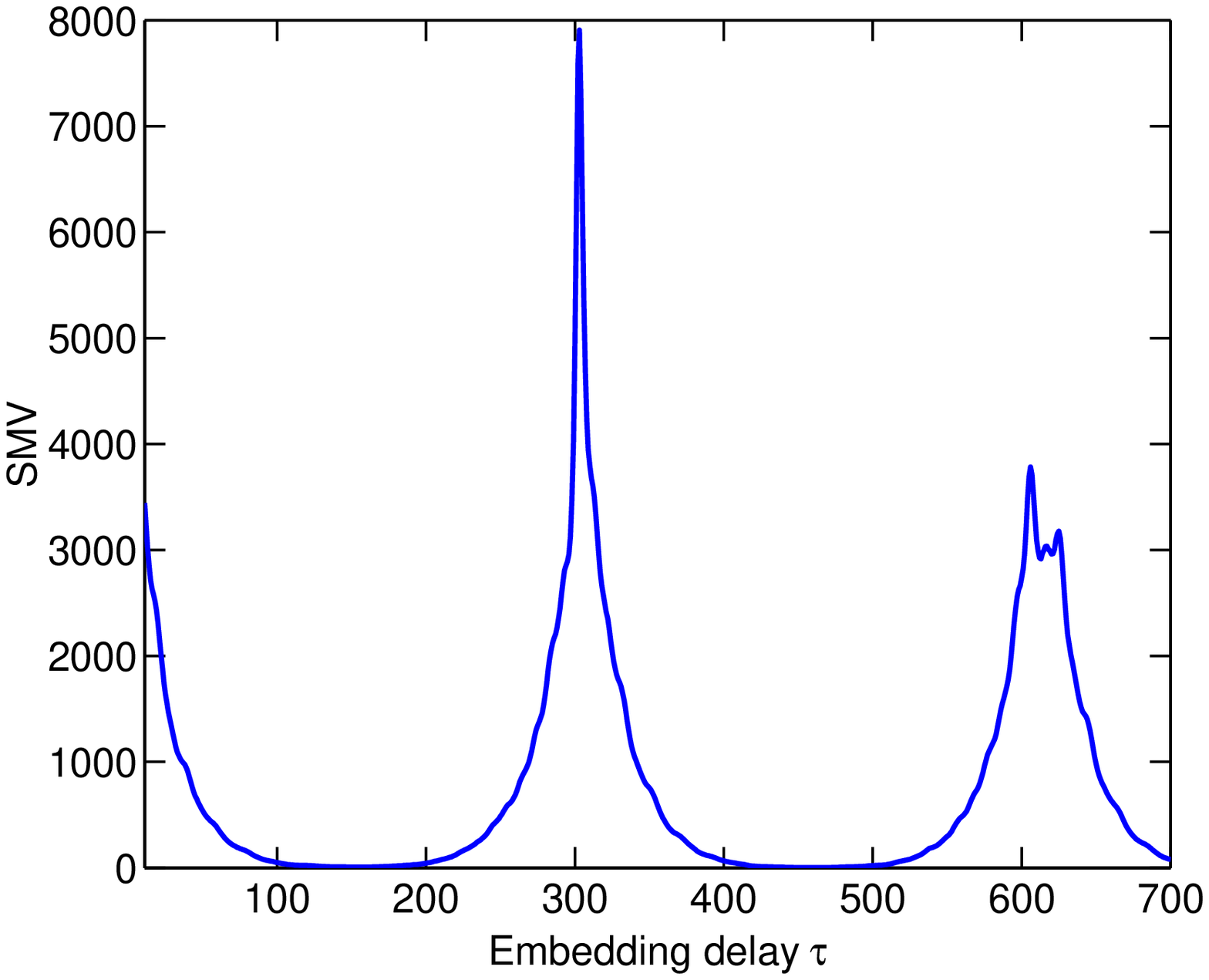}}
\subfigure[\label{fig8d}]{\includegraphics[width=7.0cm]{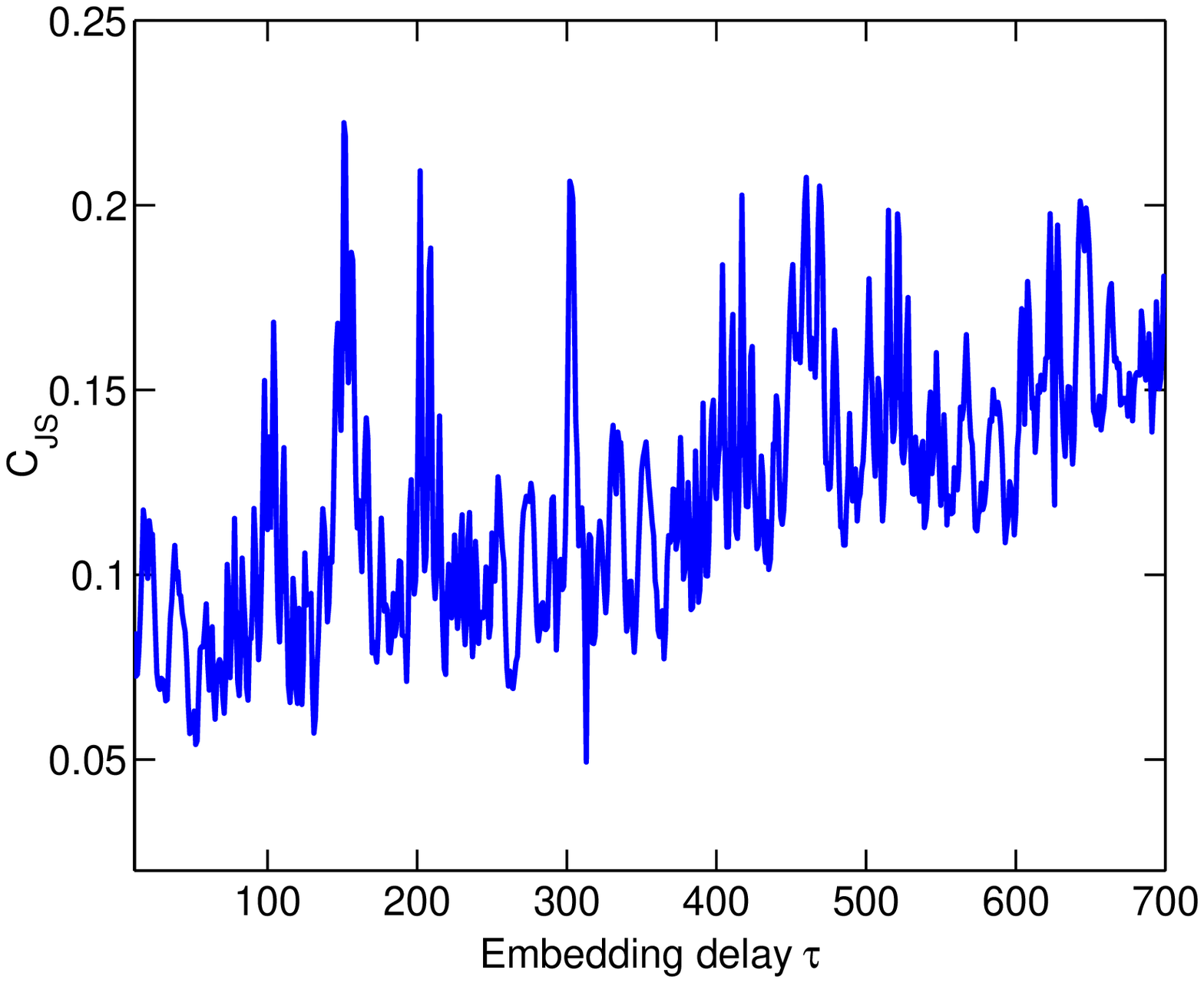}}
\caption{\label{fig8} Comparison between the SMV and the ${C_{JS}}$ for the Mackey-Glass time series with different data lengths and $NL = 0.2$: (a) and (b) $N = 15000$, and (c) and (d) $N = 10000$. The number of data segments $L$ for the proposed SMV is equal to 5. The embedding dimensions for the ${C_{JS}}$ in (b) and (d) are both 6.}
\end{figure*}
In Fig.~\ref{fig7}(a) we plot the SMV as a function of embedding delay $\tau $ of the incomplete reconstruction for different levels of observational noise, and in Fig.~\ref{fig7}(b) we plot the enlargement near the time delay ${\tau _{0}}$ of the system in order to observe more clearly. It can be observed that the proposed method is very robust in the presence of the observational noise. The computation of the SMV is based on a comparison of all the points in the first component of the incomplete two-dimensional reconstructed phase-space, and the relationship among the points of the first component can not be completely destroyed by the observational noise. Hence, the proposed SMV can recover the time delay of the system in the presence of observational noise.
\subsection{The effect of data length}
In practice, the data we need may not be always sufficient, and most of methods fail to recover the time delay ${\tau_0}$ of the system due to the data shortage. Thus, it is of significance to study the effect of data length on the proposed SMV. The effect of data length on the ${C_{JS}}$ is also studied for the purpose of comparison. The embedding dimension $m = 6$ is considered when calculating the ${C_{JS}}$ in the case of small amount of data. The observational noise with $NL = 0.2$ is added to the Mackey-Glass time series. Fig.~\ref{fig8} compares the results obtained for the SMV and the ${C_{JS}}$ for two different data lengths $N = 15000$ and $N = 10000$. From Fig.~\ref{fig8} it is found that our approach is able to recover the time delay ${\tau_0}$ of the system successfully in both situations. The amplitude of the peak associated with the time delay ${\tau_0}$ of the system just becomes lower when the amount of data is of small value. However, the ${C_{JS}}$ cannot reveal the correct time delay ${\tau_0}$ when $N=10000$, as shown in Fig.~\ref{fig8}(d). It should be noted that other embedding dimensions ($m = 3,4,5$) for ${C_{JS}}$ are also can not recover the time delay ${\tau_0}$ of the system in this situation. Actually, the ${C_{JS}}$ can not recover the time delay ${\tau_0}$ any more when the amount of data is relatively small, i.e., for $N < 10000$.
\begin{figure}[htbp!]
\includegraphics[width=7.0cm]{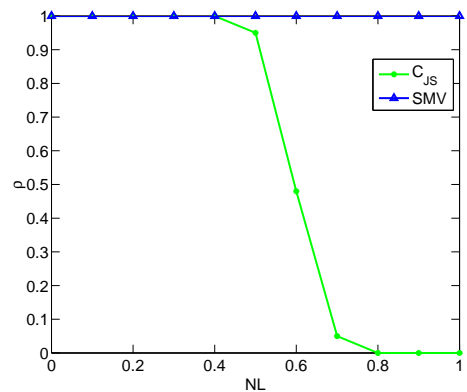}
\caption{\label{fig9} Identification rate $\rho $ as a function of $NL$ for the proposed SMV and the ${C_{JS}}$. The number of data segments $L$ for the SMV is equal to 5. The embedding dimension $m$ for the ${C_{JS}}$ is 6, and the simulation is carried out with  $100$ Monte Carlo experiments and is averaged. $N = 15000$ data are used.}
\end{figure}
\begin{figure}[htpb!]
\centering
\subfigure[\label{fig10a}]{\includegraphics[width=7.0cm]{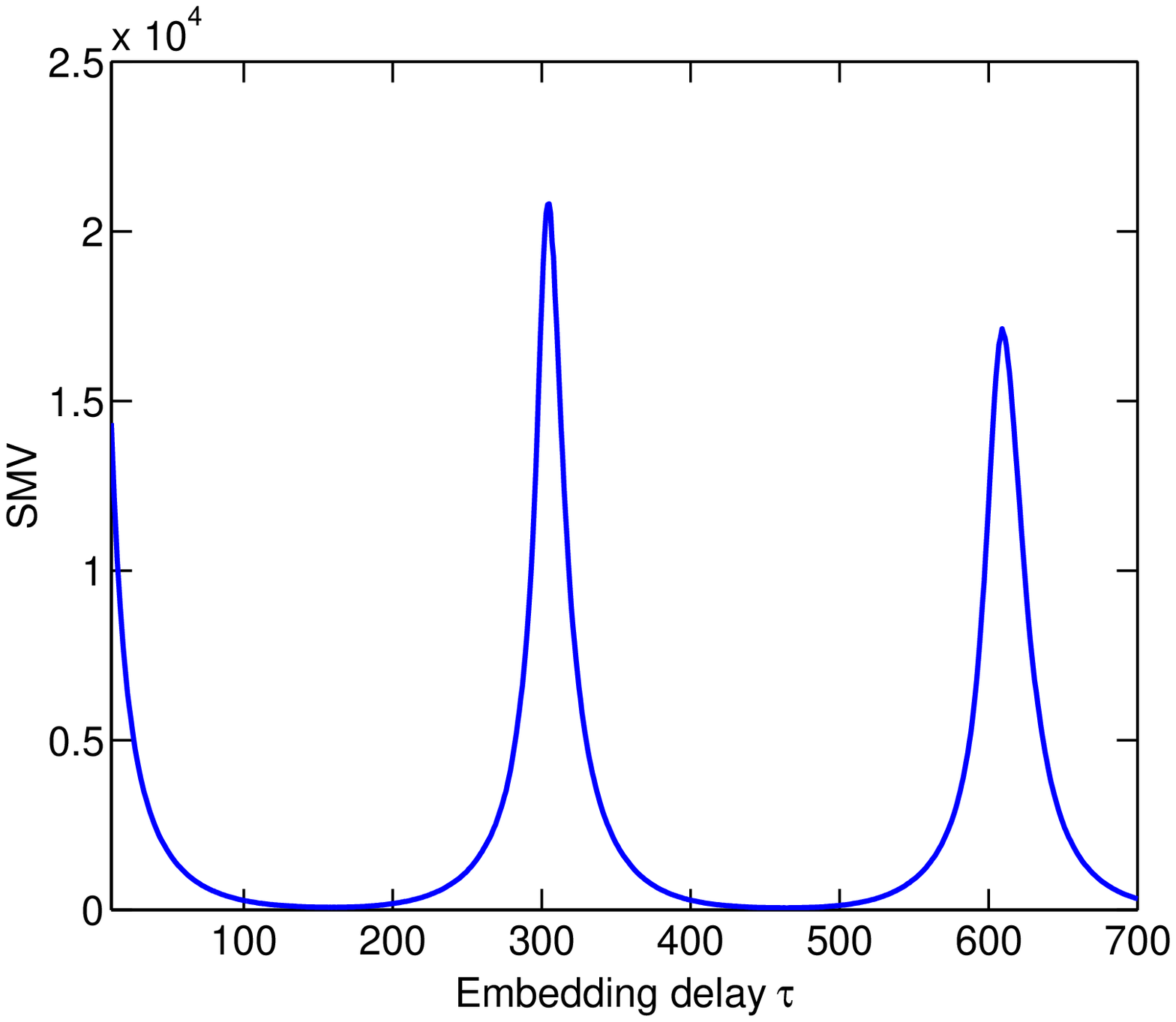}}
\subfigure[\label{fig10b}]{\includegraphics[width=7.0cm]{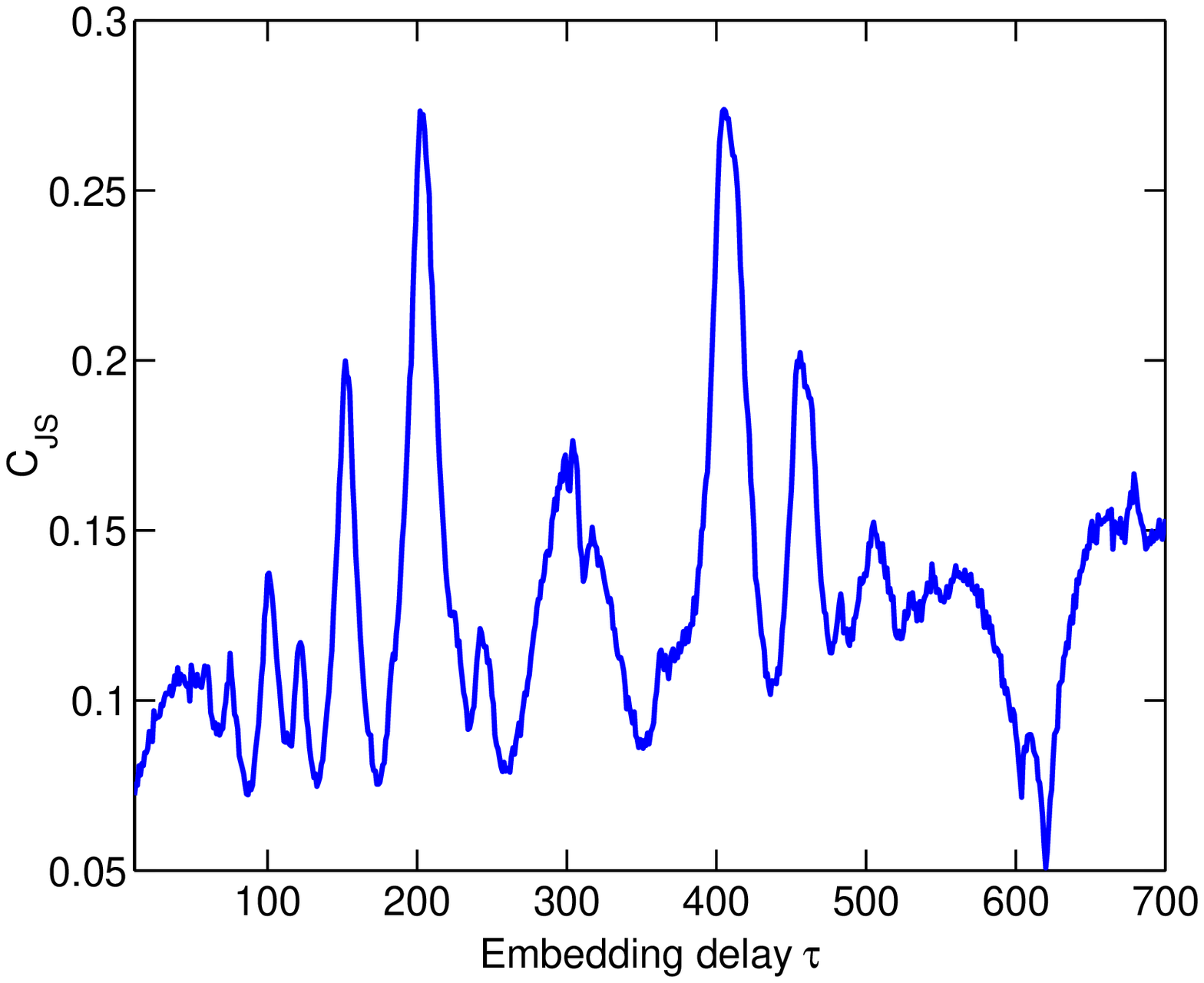}}
\caption{\label{fig10} Comparison between the SMV and the ${C_{JS}}$ with low feedback strength($\alpha = 1.4$) and noise level $ NL  = 0.2$: (a) the SMV, and (b) the ${C_{JS}}$. The number of data segments for the SMV is $L=5$, the embedding dimension for ${C_{JS}}$ is $m=6$. $N = 15000$ data points are used.}
\end{figure}

In order to better describe the effect of data length on these two methods and for the purpose of comparison, we will exploit the location of the embedding delay $\tau$ associated with the largest SMV to verify the accuracy of time delay identification in the present paper. In consideration of time delay overestimation, a vicinity $W\left( {{\tau _0}} \right)$ of the time delay ${\tau _0}$ of the system is defined as\cite{xiang2011conceal}:
\begin{eqnarray}
W\left( {{\tau _0}} \right) = \left[ {{\tau _0} - \varepsilon  \times {\tau _0},{\tau _0} + \varepsilon  \times {\tau _0}} \right],
\end{eqnarray}
where $\varepsilon $ is the mismatch coefficient, and $\varepsilon $ is set to ${\rm{5}}\% $ in this paper. The time delay is still considered to be estimated successfully once the embedding delay $\tau$ associated with the largest SMV locates in the vicinity $W\left( {{\tau _0}} \right)$. The identification rate $\rho $ is defined as the ratio of the number of successful identifications and the number of trials. The simulation is carried out with 100 Monte Carlo experiments and is averaged, the results are depicted in Fig.~\ref{fig9}.

We can see from Fig.~\ref{fig9} that the proposed method is able to recover the time delay of the system even though the level of observational noise is higher. By contrast, the performance of the ${C_{JS}}$ begins to decline when $NL > 0.4$, and the identification rate $\rho $ of the ${C_{JS}}$ is zero when $NL > 0.8$, which means that the ${C_{JS}}$ is failed to recover the time delay of the system in the case of small amount of data with a large amount of observational noise. It can be concluded that the performance of the SMV is much better than that of the ${C_{JS}}$ in the case of small amount of data and higher levels of observational noise.
\subsection{The effect of feedback strength}
It is well known that the recovery of the time delay would be difficult if the feedback strength is small, it was also pointed out that the identification of the time delay of the system would be impossible when the optical feedback of the chaotic semiconductor laser was weak\cite{rontani2009time}. In order to check the performance of the proposed SMV in this severe time delay identification scenario, numerical simulation of the data generated by Eq.~(\ref{equationmg}) with the same parameters (${\tau _0}{\rm{ = }}60,\gamma  = 10$) but low feedback strength $\alpha  = 1.4$, is analyzed.
Meanwhile, the ${C_{JS}}$ is also used to identify the time delay of the system for the purpose of comparison. $N=15000$ data points are used in the experiment for the reason that the ${C_{JS}}$ can not recover the time delay of the system if the data length is too small.
The results are shown in Fig.~\ref{fig10}.

We can see from Fig.~\ref{fig10}(a) that the proposed SMV shows pronounced maximum when the embedding delay $\tau$ of the system is close to the time delay of the system, which means that the proposed method can recover the time delay of the system when the feedback strength is small. Moreover, we can also observe from Fig.~\ref{fig9}(a) that the amplitude of the peak associated with $\tau  = 600$ becomes larger than that of the peak associated with $\tau  = 600$ in Fig.~\ref{fig9}(a), this is due to the effect of the low feedback strength.
As for the ${C_{JS}}$, it fails to recover the time delay in this situation, as shown in Fig.~\ref{fig10}(b).
From Fig.~\ref{fig10} we can conclude that the performance of the proposed SMV is much better than that of the ${C_{JS}}$ in the case of weak feedback strength.
\begin{table*}[hbtp!]
\caption{\label{table1}
Time consumptions of the SMV and the ${C_{JS}}$ with different data lengths (unit:s).}
\begin{ruledtabular}
\begin{tabular}{cccccc}
 &$ N= 120 $ &$N = 600$ &$N = 3600$
 &$ N= 25200$ &$ N= 201600$ \\
\hline
${C_{JS}}$& $2.9915 \times {10^{ - 4}}$ & $3.1444 \times {10^{ - 4}}$ & $6.5508 \times {10^{ - 4}}$ & 0.004532 & 0.386460 \\
SMV & $8.2859 \times {10^{ - 5}}$ & $1.1546 \times {10^{ - 4}}$ & $2.7783 \times {10^{ - 4}}$ & 0.001071 & 0.073973 \\
\end{tabular}
\end{ruledtabular}
\end{table*}
\subsection{The time complexity}
The time complexity is also an important consideration in practical applications, especially when the amount of data is relatively large. From Sec.~\ref{sec2} we know that the calculation of the SMV is quite simple since one only need to compute the mean and variance of the data. In contrast, the computation of the ${C_{JS}}$ is somewhat complicated when the embedding dimension $m$ is of large value. In order to compare the time complexity of these two approaches, the average time consumptions via $100$ Mente Carlo runs of the SMV and the ${C_{JS}}$ for different data lengths are shown in Table~\ref{table1}, the data lengths $N$ we choose here is corresponding to the embedding dimension $m = 4,5,6,7,8$ for ${C_{JS}}$ subjected to $N=5m!$. It is found that the time consumptions of the SMV are lower than that of the ${C_{JS}}$ in all situations, and the difference becomes obvious when the amount of data is of large value. The simulations are run on the computer with a $3.40GHz$ Intel Core i$7$-$2600$K CPU and an $8.00GB$ RAM. The release of MATLAB is $2012b$.

\section{\label{sec4}Conclusions}
Time-delay chaotic systems are widely used in practice because of their high degree of nonlinearity and complex dynamics. In such systems, time delay always plays a vital role and can provide additional information about the relationship between different components. The recovery of time delay present in the time series is one of the key problems in the study of time-delay chaotic systems.
However, time delay identification is not an easy task due to the shortage of a prior knowledge, the small amount of data and the effect of noise. In this paper, a computationally quick and conceptually simple approach is introduced to deal with this task. Before that, we propose an important procedure called the component permutation to show the local clustering phenomenon of the chaotic system based on the incomplete two-dimensional reconstruction of dynamics of the system.
We find that the amount of information captured by the incomplete two-dimensional reconstructed phase-space is associated with the time delay present in the time-delay chaotic system.
Furthermore, the information can be transferred to the permutation component by the procedure of component permutation.
Then, a statistic SMV is developed from the permutation component to recover the time delay present in the system. Numerical data generated by the time-delay systems based on the well-known Mackey-Glass equation are used to test the effectiveness and reliability of the proposed method. Numerical results show that the proposed SMV is robust to additive observational noise, and is able to recover the time delay of the chaotic system with small feedback strength. What's more, it is found that the performance of the proposed method is also good in the case of small amount of data contaminated by a large amount of observational noise. The time complexity of the proposed SMV is also quite low.
\begin{acknowledgments}
This work was supported by the Joint Fund of the National Natural Science Foundation of China and the China Academy of
Engineering Physics (Grant No. U1530126).
\end{acknowledgments}
\bibliography{aps_TDI}
\end{document}